\documentclass[prl,twocolumn,showpacs,amsmath,amssymb,superscriptaddress]{revtex4}

\usepackage{amsmath,amssymb} 
\usepackage{bm}
\usepackage[dvips]{graphicx}
\usepackage{epsfig}     
\usepackage{dcolumn}
\usepackage{bm}

\begin{document}

\title{Magneto-optical Trapping of Cadmium}
\author{K.-A.~Brickman}
\affiliation{FOCUS Center and Department of Physics, University of Michigan, Ann Arbor, MI  48109}
\author{M.-S.~Chang}
\affiliation{FOCUS Center and Department of Physics, University of Michigan, Ann Arbor, MI  48109}
\author{M.~Acton}
\affiliation{FOCUS Center and Department of Physics, University of Michigan, Ann Arbor, MI  48109}
\author{A.~Chew}
\affiliation{FOCUS Center and Department of Physics, University of Michigan, Ann Arbor, MI  48109}
\author{D.~Matsukevich}
\affiliation{FOCUS Center and Department of Physics, University of Michigan, Ann Arbor, MI  48109}
\author{P. C.~Haljan}
\affiliation{Physics Department, Simon Fraser University, Burnaby, BC V5A 1S6, Canada}
\author{V. S.~Bagnato}
\affiliation{Instituto de Fisica e Quimica de S\~{a}o Carlos, Caixa Postal 369, 13560 S\~{a}o Carlos, S\~{a}o Paulo, Brazil}
\author{C.~Monroe}
\affiliation{FOCUS Center and Department of Physics, University of Michigan, Ann Arbor, MI  48109}

\date{\today}
\begin{abstract}
We report the laser-cooling and confinement of Cd atoms in a magneto-optical trap, and characterize the loading process from the background Cd vapor.  The trapping laser drives the $^{1}S_{0}$~$\rightarrow$~$^{1}P_{1}$ transition at 229 nm in this two-electron atom and also photoionizes atoms directly from the $^{1}P_{1}$ state.  This photoionization overwhelms the other loss mechanisms and allows a direct measurement of the photoionization cross section, which we measure to be $2 (1) \times 10^{-16}$~cm$^{2}$ from the $^{1}P_{1}$ state.  When combined with nearby laser-cooled and trapped Cd$^{+}$ ions, this apparatus could facilitate studies in ultracold interactions between atoms and ions.

\end{abstract}
\maketitle

\section{Introduction}
The magneto-optical trap (MOT) is an indispensable source of cold atoms for a range of studies and applications in atomic physics, from precision atomic spectroscopy \cite{hoyt95} and cold collisions \cite{weiner99} to atom interferometry and the generation of quantum-degenerate gases \cite{leggett01}.  While nearly all cold atom experiments deal with the alkali atoms, there has been progress in the trapping of two-electron atomic species such as Ca, Mg, Sr, and Yb \cite{oates99,sengstock94,xu02,park03}, mainly for experiments involving high resolution spectroscopy of the $^{1}S$ $\rightarrow$ $^{3}P$ intercombination lines.  We report here the trapping of atomic Cd atoms in a deep-ultraviolet MOT operating on the $^{1}S_{0}$ $\rightarrow$ $^{1}P_{1}$ transition at 229 nm. 

When producing a Cd MOT, the trapping light can also photoionize the atoms directly from the $^{1}P_{1}$ excited state. While this introduces losses on the trapping process, it also provides an opportunity to reliably create cold ions and atoms at the same location \cite{cetina07} for the investigation of ultracold atom-ion interactions \cite{cote02,idziaszek07}.  One interesting future possibility is the transfer of coherence between ground state hyperfine levels in a trapped ion to pure nuclear spin states in a neutral atom lacking electron spin.  Because the nuclear spin can be extremely well-isolated from environmental influences \cite{chupp94}, control of such a coherent transfer process may have applications to the long term storage of quantum information.

In this work we realize the first Cd MOT and characterize the various trapping parameters.  Results are compared with simple analytic and Monte-Carlo simulation models of the trapping process. Through a detailed investigation of the loss rate as a function of laser intensity, the absolute photoionization cross section from the $^{1}P_{1}$ state is determined. 
  
\section{Background}
\begin{figure}
\begin{center}
\includegraphics[width=1.0\columnwidth,keepaspectratio]{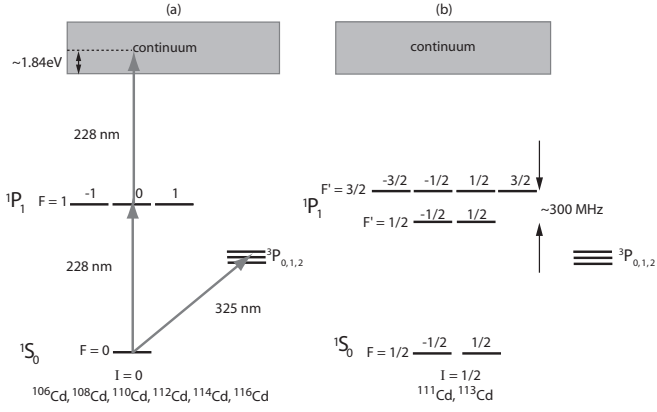} 
\caption{Cadmium energy level diagram (not to scale). a.  The bosonic (even) isotopes (I=0) of Cd.  b.  The fermionic (odd) isotopes (I=1/2) of Cd, where the $^{1}P_{1}$ hyperfine splitting arises from (L $\cdot$ I) coupling.  Individual levels are labeled with m$_{F}$.}
\label{fig:e-level}
\end{center}
\end{figure}

Cadmium has eight stable isotopes, six of which are relatively abundant.  Fig.~\ref{fig:e-level} shows the electronic structure of Cd for both bosons (nuclear spin I=0, even isotopes) and fermions (I=1/2, odd isotopes).  Most of the the data presented here is for $^{112}$Cd.  The $^{1}S_{0}$ - $^{1}P_{1}$ atomic transition used for the MOT occurs at a wavelength of $\lambda$=228.8 nm with an excited state lifetime of $\tau$=1.8 ns (radiative linewidth $\gamma/2\pi$ = 91 MHz) and saturation intensity of $I_{sat}$=$\pi h c \gamma/(3 \lambda^{3})$ $\sim$ 1.0 W/cm$^{2}$.  The saturated photon recoil acceleration on a Cd atom is ${a_{0} = \hbar /2 \tau m \lambda =4.4 \times 10^{5} g}$, which is 50 times that of Rb (here $g$ is the acceleration due to gravity and $m$ is the mass of a single Cd atom).  Note that the 228.8 nm light can also excite atoms from the $^{1}P_{1}$ state directly to the ionization continuum. 

In a vapor cell, the radiative forces accumulate atoms following the rate equation 

\begin{equation}
\label{dN/dt}
\frac{dN}{dt}  = L- \Gamma N-\beta \frac{N^{2}}{V}, \\
 \end{equation}

\noindent where $N$ is the number of trapped atoms, $L$ is the loading rate, $\Gamma$ is the loss rate related to single atom effects, $\beta$ is the loss rate due to binary collisions within the trap, and $V$ is the effective volume occupied by the trapped atoms \cite{monroe1990,hoffman94,santos95,dinneen99}.  Using simple kinetic gas theory at constant temperature one can show that $L  \approx \it{n} V_{c}^{2/3} v_{c}^{4}/v_{th}^3$, where $V_{c}$ is the capture volume, $v_{c}$ is the capture velocity \cite{monroe1990}, $v_{th}$ is the thermal velocity, and $n$ is the density of Cd atoms in the background vapor \cite{lindquist92,gibble92}.  For comparison to the data we use a simple analytic 1-D laser cooling model to find the capture velocity, as detailed in appendix 1.  Appendix 2 presents a 3-D Monte-Carlo simulation, which includes magnetic field and polarization effects, to directly estimate the loading rate.

When the MOT density is low ($<$ $10^{9}$ atoms/cm$^{3}$), the atoms are essentially non-interacting and we expect the density to be limited by the cloud temperature.  In this regime the spatial distribution of trapped atoms is expected to be Gaussian with a cloud radius that is independent of the trapped atom number.  This contrasts with high density ($>$$10^{10}$ atoms/cm$^{3}$) MOTs where effects such as reradiation \cite{walker90} must be considered.  The Cd MOT reported here operates in the low density regime, and the last term of Eq.~\ref{dN/dt} can be neglected.  Unlike conventional alkali  MOTs, where single atom loss mechanisms primarily involve collisions between trapped atoms and the background gas, Cd (like Mg) has an additional single atom loss term due to photoionization \cite{madsen2002,deslauriers2006}.  Solving Eq.~\ref{dN/dt} for the steady state number of trapped atoms gives $N_{ss} = L/\Gamma$, with the loss rate given by

\begin{eqnarray}
\label{growth time}
\Gamma & = & \Gamma_{0} + \Gamma_{ion}.
\end{eqnarray}

\noindent Here $\Gamma_{0}$ represents the rate at which trapped atoms are ejected due to collisions with the background vapor (dominated by Cd) and $\Gamma_{ion}$ is the photoionization rate:

\begin{equation}
\label{gamma ion}
\Gamma_{ion}= \frac{\sigma P(I,\delta) I}{\hbar \omega}.
 \end{equation}

\noindent In this expression, $\sigma$ is the photoionization cross section, $\hbar \omega$ is the photon energy, $I$ is the total MOT laser beam intensity, and P(I, $\delta$) is the fraction of atoms in the excited state ($^{1}P_{1}$) defined as

\begin{equation}
\label{excited state pop}
P(I, \delta)= \frac{s}{2(1+s+4\delta^{2})},
 \end{equation}

\noindent where $\delta=\Delta/\gamma$ is the laser detuning scaled to the natural linewidth and $s=I/I_{sat}$ is the saturation parameter. 

\section{Experimental Set-up and Procedure}
\begin{figure}
\begin{center}
\includegraphics[width=1.0\columnwidth,keepaspectratio]{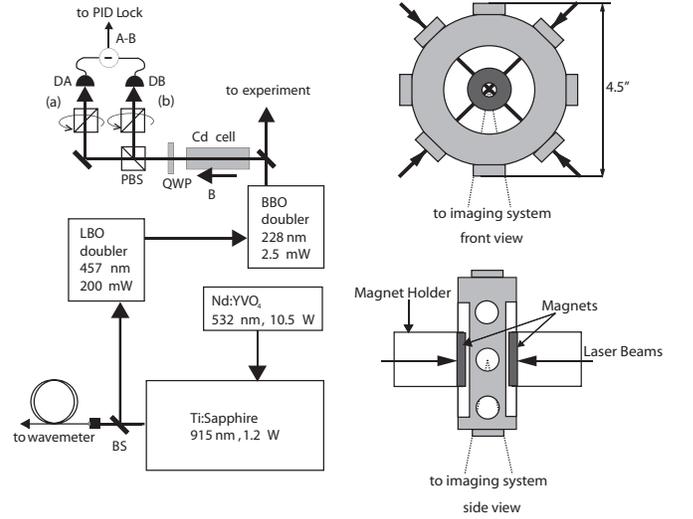} 
\caption{Left: Schematic diagram of the laser system and the laser lock (DAVLL).  The laser lock consists of the Cd cell, a quarter wave plate (QWP), a polarizing beam splitter (PBS), and two photodetectors (DA, DB) for path a and b, respectively.  Right: The MOT vacuum chamber and the laser beam geometry.  The MOT is formed by 6 independent beams.  The imaging system sits below the chamber, and the dark shaded regions are the NdFeB magnets.}
\label{fig:table-setup}
\end{center}
\end{figure}

A schematic of the experimental apparatus is shown in Fig.~\ref{fig:table-setup}.  Since Cd has a large linewidth, high magnetic field gradients are required to shift the Zeeman levels sufficiently for the atoms to feel a substantial trapping force at the edge of the laser beams.  We use NdFeB permanent ring magnets with a 2.54 cm outer diameter, 0.64 cm inner diameter, and 0.95 cm thickness that are mounted coaxially on translational stages.  By adjusting the axial separation of the magnets we can achieve magnetic field gradients up to 1500 G/cm at the trap center.

The trapping beams are generated with a frequency quadrupled Ti:Sapphire laser, yielding 2.5 mW at 228.8 nm.  The ultraviolet light is split into six independent trapping beams in order to better control the intensity balance of the counter-propagating beams.  The MOT can withstand an intensity imbalance of 10$\%$ between a pair of beams, and we can balance the intensity between any pair of counter-propagating beams to better than 5$\%$.  Typical beam waists range from $w$ =0.5~mm to 1.5~mm and the total power ranges from $P$= 0.7 mW to 2.0 mW, resulting in peak intensities ranging from about 0.03 W/cm$^{2}$ to 0.5 W/cm$^{2}$.

Approximately 200 $\mu$W is split from the main laser beam and directed to a small cadmium vapor cell to stabilize the laser frequency.  We use a dichroic atomic vapor laser lock (DAVLL) \cite{corwin98,reeves06} operating on the $^{1}S_{0}$ $\rightarrow$ $^{1}P_{1}$ transition in Cd.  The cell is heated to $80^{\circ}$ C to increase optical absorption to about $80\%$ through the 5 cm cell.  A uniform magnetic field is applied along the laser beam axis to lift the degeneracy of the $^{1}P_{1}$ states.  When linearly polarized light is sent through the cell the difference between absorption of the Zeeman-shifted $\sigma^{+}$ and $\sigma^{-}$ transitions produces a dispersive-shaped signal and the laser is locked to the zero crossing point of this signal.  The capture range is determined by the Zeeman splitting between the two transitions, or about 1.5 GHz in a 500 G field. To change the detuning we move the zero crossing point by attenuating the laser power in one of the polarization paths (a or b) after the cell (see Fig.~\ref{fig:table-setup}). The lock is stable to within 30 MHz, or $0.3\gamma$, over the 1.5 GHz capture range, and the dominant sources of fluctuations are beam-steering drifts and birefringence fluctuations of the cell windows from temperature drifts over times greater than 1 second.

The MOT chamber contains a 1 cm long hollow stainless steel tube of diameter 0.1 cm packed with about 0.02 g of pure Cd wire.  We control the background Cd vapor pressure throughout the entire chamber by heating this small oven.  When we direct the trapping beams into the chamber we see tracks of fluorescing Cd within the extent of the laser beams. Based on this atomic fluorescence, we estimate the background Cd vapor pressure to range between approximately 10$^{-11}$ torr with the oven off to about 10$^{-10}$ torr with the oven at approximately $300^{\circ}$~C.  We speculate that the Cd atoms sublimated from the oven do not readily stick to the chamber surface, resulting in good control of the Cd vapor pressure with the small oven.   We note that the vapor pressure of Cd is predicted to be 10$^{-11}$ torr at room temperature \cite{Adzhimambetov03}, which is consistent with our observations.

The atomic fluorescence from the trapped atoms is collected with an f/3 lens (a solid angle of d$\Omega$/4$\pi$ = 0.6$\%$) and imaged onto an intensified charge coupled device (ICCD) camera.  Every photon incident on the camera yields $\eta$G $\simeq$ 65 counts, where $\eta$=20$\%$ is the quantum efficiency of the camera and $G$ is the ICCD gain factor.  Including an optical transmission of $T\sim50\%$ in the imaging system, we expect a total count rate of $\gamma P(I,\delta)G\eta T(d\Omega/4\pi) \sim 10^{7}$~counts/sec from each trapped atom in the MOT.  In this way, we relate the total fluorescence count rate to the number of atoms in the MOT, with an estimated accuracy of 50$\%$.    For various settings of the MOT parameters, we are able to observe between about 10-3000 atoms in the MOT.

\section{Results and Discussion}
\begin{figure}
\begin{center}
\includegraphics[width=1.0\columnwidth,keepaspectratio]{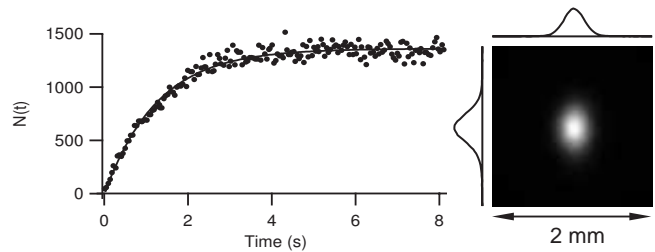} 
\caption{Left: Typical loading curve showing the buildup in the MOT fluorescence vs.~time.  For this data set, the MOT parameters are laser power $P$=1.45~mW, beam waist $w$=1.25~mm, detuning $\delta$=$-$0.7, and magnetic field gradient B$^{\prime}$=500~G/cm.  The steady state MOT number is calculated from the fluorescence signal and for this data the buildup time is 1.5 sec.  Right: MOT image taken with the camera for N$_{ss}$=1200~atoms.  The MOT parameters for this data set are $P=$1.45~mW, $w$=2.5~mm, $\delta$=$-$0.7, and B$^{\prime}$=500~G/cm.  The integration time for the camera was 5 ms.  A 2-D Gaussian fit to the image yields an rms radius of 200~$\mu$m and a peak atom density of 10$^{8}$~ atoms/cm$^{3}$.}
\label{fig:loadingcurve}
\end{center}
\end{figure}

A typical observation of the fluorescence growth from trapped atoms in time is shown in Fig.~\ref{fig:loadingcurve}, allowing a determination of the steady-state number of atoms and the net loss rate, $\Gamma$, from the trap.  An image of the fluorescence distribution from the trapped atoms is also shown in Fig.~\ref{fig:loadingcurve}, revealing a Gaussian-shaped atom cloud as expected from the temperature-limited density.  The typical geometric mean rms radius of the MOT is 200 $\mu$m, with some dependence upon the magnetic field gradient, laser power and detuning. The largest MOT we have observed held approximately $3000$ atoms, with a peak density of about 10$^{8}$~atoms/cm$^{3}$.

\begin{figure}[h!]
\begin{center}
\includegraphics[width=0.8\columnwidth,keepaspectratio]{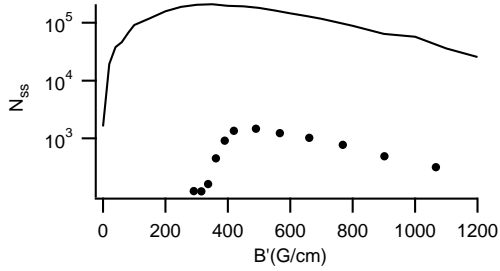} 
\caption{Observed steady-state MOT number vs.~axial magnetic field gradient $B^{\prime}$(points), along with the 3-D model (solid line) for $P$=0.8~mW, $\delta$=$-$0.6, and $w$=2.5~mm.}
\label{fig:numvsb}
\end{center}
\end{figure}

Fig.~\ref{fig:numvsb} shows the steady state number of atoms, $N_{ss}$, in the MOT vs.~magnetic field gradient, $B^{\prime}$, for beam waist $w$=1.25 mm, detuning $\delta$=$-$0.6, and a total power $P$=1.8 mW.  Under these conditions the maximum steady state number is observed at 500 G/cm.  At this optimal field gradient, the Zeeman shift of the excited state levels at the edge of the laser beam is approximately one linewidth.  Above this optimal value the steep magnetic field gradient shifts the atoms out of resonance with the laser beams, reducing the capture volume.  At lower field gradients N$_{ss}$ quickly decreases, presumably due to a lower trap depth resulting from an increased sensitivity to trapping parameters.

From the equipartition theorem we obtain a relation connecting the cloud radius and temperature, $\kappa r^{2} = k_{b}T$, where $r$ is the atomic cloud rms radius, $k_{b}$ is Boltzmann's constant, $T$ is the temperature in Kelvin, and $\kappa$ is the trap spring constant $\kappa=8\mu_{b}ksB^{\prime}\delta/(1+s+4\delta^{2})$ \cite{metcalf}.  In this expression, $\mu_{b}$ is the Bohr magneton, and $k = 2\pi/\lambda$ is the wavenumber.  Replacing $T$ with the Doppler temperature, $T_{D} =\hbar\gamma (1+s+4\delta^2)/(8k_{b}|\delta|)$, gives a relation between the temperature-limited cloud radius and the magnetic field gradient:

\begin{equation}
\label{motsize}
r = \sqrt{\frac{\hbar\gamma(1+s+4\delta^{2})^{3}}{64\mu_{b}\delta^{2}ksB^{\prime}}}.
\end{equation}

\begin{figure}[t!]
\begin{center}
\includegraphics[width=0.8\columnwidth,keepaspectratio]{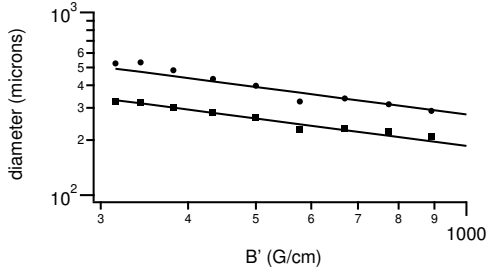} 
\caption{Atom cloud rms diameter vs.~$B^{\prime}$ for $P$=0.8~mW, $\delta$=$-$0.6, and $w$=2.5~mm.  A characterization is provided by the longest (circles) and shortest (squares) rms size of the elliptical MOT.  The diameter is about 5 times larger than what Doppler theory predicts.  The solid lines show the $(B^{\prime})^{-1/2}$ dependence expected from Eq.~\ref{motsize}.}
\label{fig:atomiccloud}
\end{center}
\end{figure}

Fig.~\ref{fig:atomiccloud} shows the MOT rms radius vs.~magnetic field gradient; as expected from Eq.~\ref{motsize}, the cloud gets smaller as B$^{\prime}$ increases.  The MOT diameter is roughly 5 times larger than what Doppler theory predicts.  Similar results were found in Sr, where the MOT temperature exceeded the expected Doppler temperature \cite{xu02}.  

\begin{figure}[h!]
\begin{center}
\includegraphics[width=0.8\columnwidth,keepaspectratio]{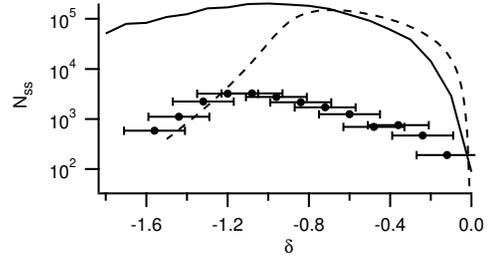} 
\caption{Observed steady-state atom number vs.~$\delta$ (points) along with the 1-D (dotted line) and 3-D (solid line) models for $P$=1.8~mW, $B^{\prime}$=500~G/cm and $w$=2.5~mm.}
\label{fig:numvsd}
\end{center}
\end{figure}

\begin{figure}[h!]
\begin{center}
\includegraphics[width=0.8\columnwidth,keepaspectratio]{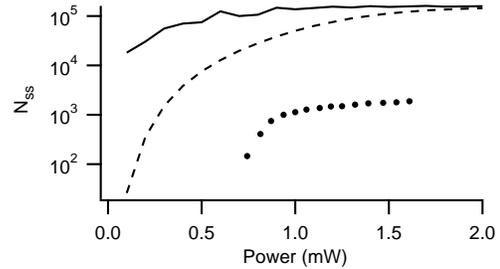} 
\caption{Observed steady-state atom number vs.~power (points) for $\delta$=$-$0.7, $B^{\prime}$=500~G/cm and $w$=2.5~mm along with the 1-D (solid line) and 3-D (dotted line) models.}
\label{fig:numvsp}
\end{center}
\end{figure}

\begin{figure}[h!]
\begin{center}
\includegraphics[width=0.8\columnwidth,keepaspectratio]{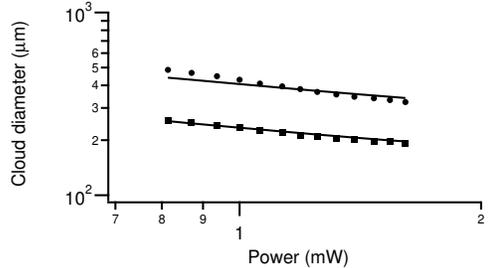} 
\caption{MOT cloud diameter vs.~total MOT laser power for $\delta$=$-$0.6, $B^{\prime}$=500~G/cm and $w$=2.5~mm.  The solid lines show the expected dependence of the MOT diameter on power from Eq.~\ref{motsize}.}
\label{fig:cloudsizevsp}
\end{center}
\end{figure}

The dependence of the steady-state number of trapped atoms on MOT detuning and laser power is shown in Figs.~\ref{fig:numvsd} and \ref{fig:numvsp}.  In both figures, the experimental data is plotted along with the 1-D and 3-D theoretical predictions.  The observed number of trapped atoms is 1-2 orders of magnitude below predictions, likely due to alignment imperfections and intensity imbalances not included in the models.  Fig.~\ref{fig:cloudsizevsp}
shows how the measured atom cloud size decreases as the MOT laser power is increased (at a fixed beam waist), as expected from Eq.~\ref{motsize}.

\begin{figure}[h!]
\begin{center}
\includegraphics[width=0.8\columnwidth,keepaspectratio]{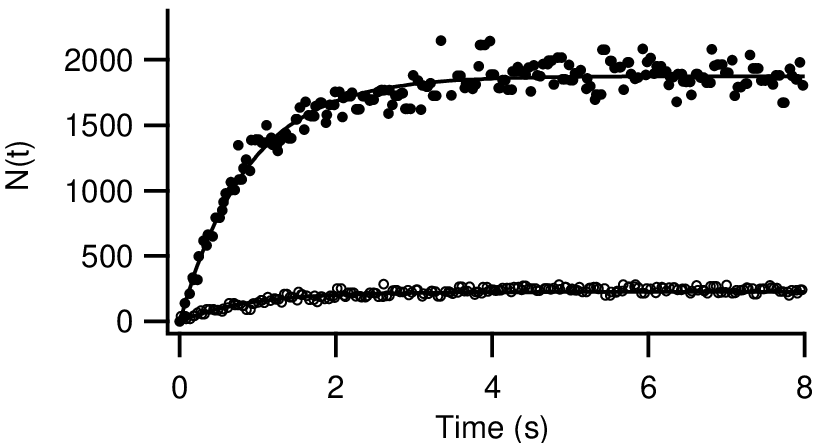}
\includegraphics[width=0.8\columnwidth,keepaspectratio]{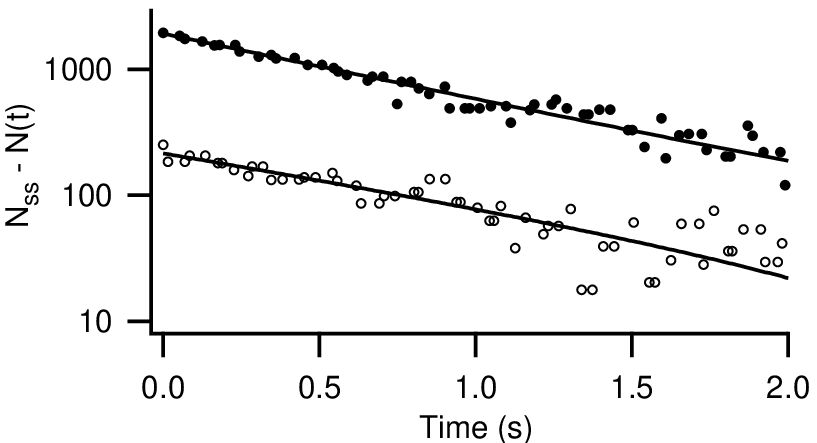} 
 \caption{Top: Observed trapped atom number N(t) for two different Cd background vapor pressures.  The top curve corresponds to a pressure of $10^{-10}$~torr and the lower curve corresponds to $10^{-11}$~torr.  By fitting the data to a growing exponential, $N(t)=N_{ss}(1-e^{-\Gamma t})$, we find that the filling time, $\Gamma^{-1}$, is approximately 1 sec for each case.  This is clear from the lower logarithmic plot of the data.  Bottom:  N$_{ss}$-N(t) plotted for both vapor pressures on a log scale.  The filling times are about 1 sec for each curve.}
\label{loadtimenooven}
\end{center}
\end{figure}

\begin{figure}[h!]
\begin{center}
\includegraphics[width=0.8\columnwidth,keepaspectratio]{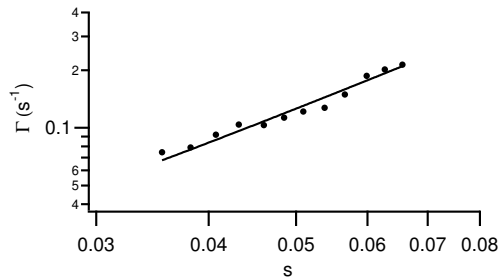} 
\caption{Observed loading rate vs the saturation parameter s=I/I$_{sat}$.  The power is varied for a constant beam waist of $w$=1.25~mm. The photoionization cross section out of the $^{1}P_{1}$ state is determined from a quadratic fit to $s$ given by Eq.~\ref{growth time}.  Extrapolating the curve to zero intensity (not shown here) gives information on the loss rate due to collisions with background gas.}
\label{cross-section}
\end{center}
\end{figure}

In Fig.~\ref{loadtimenooven}, the filling of the MOT is shown for Cd vapor pressures of approximately 10$^{-10}$ torr and 10$^{-11}$ torr.  Unlike conventional vapor cell MOTs, we find that the filling time (loss rate) is independent of the background pressure, while the steady-state number of atoms in the MOT is strongly dependent on pressure.  This indicates that collisions with the background gas have very little effect on the loss rate and instead we are limited by photoionization loss from the MOT beams.  This is investigated in more detail by measuring the filling time (loss rate) as the MOT laser intensity is varied, as shown in Fig.~\ref{cross-section}.  We observe a roughly quadratic dependence of loss rate on intensity, consistent with Eq.~\ref{growth time}.  The extrapolated loss rate at zero intensity is much smaller than all of the observations, directly indicating that $\Gamma_{0} \ll \Gamma_{ion}$, or that the loss rate in this experiment is dominated by photoionization.  From this measurement, we can also directly extract the photoionization cross section from the $^{1}P_{1}$ state, given measurements of the intensity, excited state fraction P(I,$\delta$), and the known wavelength of the light.  We find that the photoionization cross section of the $^{1}P_{1}$ state of Cd from the 228.8 nm light is $\sigma=2 (1)\times10^{-16}$~cm$^{2}$, with the error dominated by uncertainties in the laser intensity and detuning.

\begin{figure}[h!]
\begin{center}
\includegraphics[width=0.8\columnwidth,keepaspectratio]{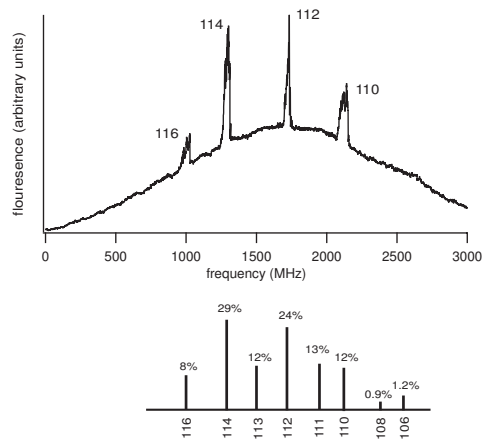} 
\caption{Top: Scan across frequency showing the different Cd isotope MOTs.  The underlying curve is the Doppler fluorescence profile of the Cd atoms.  At certain frequencies there is a large build up, due to the MOT accumulation as its resonance is crossed.  Bottom:  Natural abundance of neutral cadmium isotopes.  Out of these eight isotopes, we are only able to clearly observe trapping of the four most abundant bosonic (even) isotopes.}
\label{fig:MOT isotopes}
\end{center}
\end{figure}

\section{Fermionic Isotopes}
Scanning the laser frequency allows cooling and trapping of different cadmium isotopes, as shown in Fig.~\ref{fig:MOT isotopes}.  We observe that the peak heights correspond to the natural abundance of each isotope, showing that the bosonic isotopes are equally capable of being trapped.  However, there is a lack of evidence for the fermionic isotopes being loaded in the MOT.  This is due to the hyperfine structure present in the $^{1}P_{1}$ states of the fermionic isotopes. As shown in Fig.~\ref{fig:e-level}b, the two excited hyperfine states for both $^{111}$Cd and $^{113}$Cd are separated by about 300 MHz, which is comparable to the natural linewidth of Cd.  A laser tuned to the red of the upper hyperfine state (F$^{\prime}$=3/2) but to the blue of the lower hyperfine state (F$^{\prime}$=1/2) may drive excessive transitions to the lower excited state, which could result in too much heating and prevent trapping.  In addition, the optical transitions between the F=1/2 ground states and F$^{\prime}$=1/2 excited states do not result in spatially dependent
differential optical pumping by $\sigma ^+$ and $\sigma ^-$ transitions, a necessary condition for a standard MOT.  Similar results were reported for Yb \cite{park03}, where much smaller or no MOT was observed for fermionic isotopes.  In the present case it could be that there is a very small fermionic MOT being formed but it is not resolvable from the background noise.  It may be possible 
to laser cool and trap fermionic isotopes with a dichroic MOT \cite{flemming95}.  Here, the cooling laser is tuned to the red of the lower hyperfine transitions (F$^{\prime}$=1/2) to provide the major scattering force for laser cooling and then a small fraction of the laser power is
frequency shifted to the red of the upper hyperfine state (blue of the lower state).  When this second laser beam is collimated with a smaller beam waist, and overlapped with the beam of the first color,
the laser cooled atoms can be trapped in the MOT by driving the upper transitions (F=1/2 to F$^{\prime}$=3/2).  Alternatively, one can work in a much higher magnetic field gradient to overwhelm the excited state hyperfine structure.  In this Paschen-Bach regime, one will drive $J=0$ to $J=1$ transitions to produce a MOT. Given a beam waist of 1.0 mm, the required
field gradient for the MOT will be on the order of 10$^{4}$ G/cm, which can be realized by a pair of needle electromagnets \cite{vuletic96}.  The capture volume of the MOT will be much smaller, but this scheme may still be useful for single-atom MOT experiments.  Another alternative is to use a higher laser power allowing one to tune to the red of both hyperfine states.  With a larger detuning, $|\delta| \gg \delta_{hf}$, the optical excitation to the lower and upper manifolds is driven more evenly and can produce both cooling and trapping forces for the atoms.

\section{Conclusion}
In this paper we present the first Cd magneto-optical trap, operating on the $^{1}S_{0}$ - $^{1}P_{1}$ transition at 228.8 nm.  A characterization of the MOT as a function of magnetic field gradient, detuning, and intensity is presented.  The same beams that form the MOT also photoionize the atoms inside the MOT.  We observe photoionization as the dominant loss mechanism, and characterize the photoionization cross section.

This system, when combined with cold ions, opens the possibility of studying ultra-cold charge exchange collisions.  One outcome of these studies is the possible transfer of coherent information from the ion to the neutral atom.  A possible experiment is to prepare the ion in a quantum superposition of the hyperfine qubit states and then allow the ion to undergo an ultracold charge exchange with a nearby neutral atom.  This results in the charge neutralization of the ion, but could also leave some of the previously prepared quantum information intact in the nucleus.  This could allow quantum information to be carried by pure nuclear spins with very little interaction with the environment.  Subsequent coherent charge exchange with another ion could then allow the nuclear quantum information to be manipulated and processed using conventional ion trap techniques.  In addition to applications for quantum information, the long-lived $^{3}P_{0}$ state could be of interest for optical clocks \cite{hoyt95} and the narrow linewidth of the $^{1}S_{0}$-$^{3}P_{1}$ transition (70 kHz)  would allow for an extremely low cooling limit \cite{xu02}.

\section{Appendix}

\section{1-D derivation for  steady-state number of atoms cooled to rest in a vapor cell}

The following appendix estimates the number of atoms cooled to rest in a vapor cell.  For simplicity we assume the laser beams to have a top-hat profile.  

The force on an atom moving with velocity, $v$, in two counter-propagating laser beams is
\begin{equation}
\label{scatteringforce2}
F_{scat}=\frac{\hbar k \gamma}{2}\left[\frac{s}{1+s+4(\delta-u)^{2}}-\frac{s}{1+s+4(\delta+u)^{2}}\right],
\end{equation}

\noindent where the scaled velocity is defined as $u$=$kv/\gamma$.

To find the capture velocity $v_{c}$, that is the maximum velocity an atom can possess and still be slowed to rest within the cooling laser beams, we must integrate the velocity-dependent acceleration $a(v)$

\begin{equation}
\label{acceleration2}
\int_{v_{c}}^{0} \frac{v dv}{a(v)}=\int_{x_{0}}^{x_{0}+l} dx.
\end{equation}

\noindent Here $l$ is the laser beam diameter, $x_{o}$ is defined to be the edge of the laser beam, and 

\begin{equation}
\label{acc} 
a(v)= \frac{\hbar k \gamma}{2 m}\left[\frac{s}{1+s+4(\delta-u)^{2}}-\frac{s}{1+s+4(\delta+u)^{2}}\right].
\end{equation}

Solving Eq.~\ref{acceleration2} gives

\begin{equation}
\label{capvelfinal}
\frac{-16\hbar k^{3} s \delta l }{2m\gamma}  = \frac{16}{5} u_{c}^{5}+\frac{8}{3}(1+s -4\delta^{2}) u_{c}^{3} +(1+s+4\delta^{2})^{2} u_{c}.
\end{equation}

\noindent This fifth order polynomial can be solved numerically to find the capture velocity, $v_{c} = \gamma u_{c}/k$.

From the steady state solution to Eq.~\ref{dN/dt} given above, we get

\begin{equation}
\label{Nss}
N_{ss}=\frac{f n \hbar \omega l^{2} v_{c}^4}{v_{th}^{4} \hbar \omega \sigma_{c} n+v_{th}^3 \sigma_{ion} I P(I, \delta)},
\end{equation}

\noindent where $\sigma_{c}$ is the collision cross section, $\sigma_{ion}$ is the photoionization cross section, and $f$ is the relative abundance of the isotope of interest.

\section{3-D derivation for  steady-state number in a vapor cell MOT}

A 3-dimensional computer simulation that includes the effects of the magnetic field to obtain the steady-state number of atoms trapped in an even-isotope Cd MOT is employed to compare with our data.  Previous work has used models that calculate the capture velocity for an individual atom and extract the loading rate from $v_{c}$ \cite{aucouturier96,chevrollier96,ritchie94,gensomer97} or has included individual photon recoil events \cite{kohel03}.

Here, we treat the atoms as non-interacting, point particles and examine the dynamics of an ensemble of individual atoms under the application of the laser radiation. The atoms are subjected to a time-averaged force, i.e. we do not track individual photon absorption and re-emission events and instead we calculate the averaged momentum kicks over hundreds of scattering events.  The atoms' motional behavior is found by numerically integrating the position- and velocity-dependent radiation force.

Specifically, taking the $\hat{x}$-direction as an example, the net acceleration of an individual atom is given by the sum of the acceleration due to the $+\hat{x}$ and $-\hat{x}$ laser beams: 
\begin{eqnarray} 
\label{accelx}
a_x(\vec r, v_x) &=& \frac{\hbar k \gamma}{2m} s_x(\vec r) \bigg[ \sum_q \frac{p_{+,q}(\vec r)}{1+s_{\mathrm{tot}}(\vec r) + (2\delta_{+,q}(\vec r,v_x))^2} \nonumber \\
 & & - \sum_q \frac{p_{-,q}(\vec r)}{1+s_{\mathrm{tot}}(\vec r) + (2\delta_{-,q}(\vec r,v_x))^2} \bigg] 
\end{eqnarray}
where $q$ is the polarization index, and the $\pm$ subscripts correspond to the $+\hat{x}$- and $-\hat{x}$-direction beams, respectively. 

The individual beam saturation parameter, $s_x$, and the total saturation parameter, $s_{\mathrm{tot}}$, are given by:
\begin{eqnarray}
s_x(\vec r) &=& \frac{I_x}{I_{\mathrm{sat}}} e^{\frac{-2(y^2 + z^2)}{w_0^2}}, \\ 
s_{\mathrm{tot}}(\vec r) &=& 2 s_x(\vec r) + 2 s_y(\vec r) + 2 s_z(\vec r),
\end{eqnarray}
where $I_x$ is the intensity of the $\pm \hat{x}$ laser beams (here assumed to be balanced) and $s_y(\vec r)$ and $s_z(\vec r)$ are defined analogously to $s_x(\vec r)$. 

The fraction of the incoming laser radiation that the atom experiences as $\sigma^{\pm}$- or $\pi$-polarized is:
\begin{equation}
p_{\pm,q}(\vec r) = \left \{ \begin{array}{lll}
(\frac{1}{2}[1\mp \frac{1}{2}\frac{x B^{\prime}}{B(\vec r)}])^2, & q=-1 & (\sigma^-) \\ 
(\frac{1}{2}[1\pm \frac{1}{2}\frac{x B^{\prime}}{B(\vec r)}])^2, & q=+1 & (\sigma^+) \\
1-(p_{\pm,-1} + p_{\pm,+1}), & q=0 & (\pi)
\end{array} \right.
\end{equation}
where $B(\vec r) = B' \sqrt{z^2 +\frac{1}{4}(x^2 + y^2)}$ is the magnitude of the magnetic field written in terms of the magnetic field gradient, $B'$, along the strong ($\hat z$) axis. 

The effective detuning for the atomic transition is given by:
\begin{equation}
\delta_{\pm,q}(\vec r, v_x) = (\Delta \mp k v_x)/\gamma + q \frac{\mu_B g_F B(\vec r)}{\gamma\hbar}
\end{equation}
where $g_F=1$ is the Land\'e g-factor. Note that for the even isotopes of Cd, there is no Zeeman shift for $\pi$-polarized radiation. 

Atoms are initially placed uniformly distributed in position within a simulation volume with a lateral dimension of $6w$ (3 beam diameters) and the total number of atoms is chosen to correspond to the background density of the atomic vapor. The atoms are given initial velocities distributed according to a Maxwell-Boltzmann velocity distribution centered about 0. To save computation time, the atoms are discarded if they have an initial velocity $v_i > 5 v_c$, where $v_c \approx 20$m/s is the maximum capture velocity for the atoms calculated from the 1-dimensional model (appendix A). For select parameter sets, we have checked this time-saving assumption against a version that simulates all atoms regardless of velocity and found no difference in the obtained results. 

For each time step, $dt$, the time-averaged acceleration on each atom is computed (Eq.~\ref{accelx}) and the new velocity and position are calculated according to:
\begin{eqnarray}
v_x &=& v_x + a_x dt \\
x &=& x + v_x dt + \frac{1}{2} a_x (dt)^2.
\end{eqnarray}
At each time step, any atoms that have left the simulation volume are eliminated. The number of new atoms added at each time step is calculated by considering the average number of atoms that would leave the volume at each time step {\em assuming no radiation forces were present}.  These new atoms are added with uniformly distributed positions along one of the (randomly) chosen simulation box edges and are given velocities that point inwards but correspond to a Maxwell-Boltzmann distribution (again discarding velocities greater than $5 v_c$). 

Every 100 time steps, we count the number of atoms with positions that are within $w_0 / 2$ of the origin. As a function of time, this sum produces a linear curve with a slope that corresponds to the loading rate of atoms being trapped by the MOT.  The steady-state number of atoms confined by the MOT can then be calculated by including isotope abundance and the background collision and photoionization loss rates (similarly to Eq.~\ref{Nss}).

\end{document}